\documentclass[preprintnumbers,floatfix,letterpaper,superscriptaddress,nofootinbib,onecolumn]{revtex4-2}
\usepackage{amsmath}
\usepackage{amssymb}
\usepackage{bbm}
\usepackage{subfigure}
\usepackage{hyperref}
\usepackage{url}
\usepackage[dvipdfmx]{graphicx}
\usepackage{xcolor}
\usepackage{color}
\usepackage{mathrsfs}
\usepackage{amsfonts}
\usepackage{latexsym}
\usepackage{ragged2e}
\usepackage{epsfig}
\usepackage{textcomp}
\usepackage{phaistos}
\usepackage[title]{appendix}
\usepackage{chngcntr}

\makeatletter
\renewcommand\@makefnmark{\hbox{\@textsuperscript{\normalfont\color{purple}\@thefnmark}}}
\renewcommand\@makefntext[1]{%
	\parindent 1em\noindent
	\hb@xt@1.8em{%
		\hss\@textsuperscript{\normalfont\@thefnmark}}#1}
\makeatother

\usepackage{caption}
\DeclareCaptionJustification{justified}{\leftskip=0pt \rightskip=0pt \parfillskip=0pt plus 1fil}
\graphicspath{{./Figures2/}}

\def\beq{\begin{equation}}
	\def\eeq{\end{equation}}

\def\mathbb{\Bbb}

\definecolor{vividviolet}{rgb}{0.62, 0.0, 1.0}
\definecolor{amaranth}{rgb}{0.9, 0.17, 0.31}
\definecolor{palatinateblue}{rgb}{0.15, 0.23, 0.89}
\definecolor{brightpink}{rgb}{1.0, 0.0, 0.5}
\definecolor{cornflowerblue}{rgb}{0.39, 0.58, 0.93}
\definecolor{deepcarminepink}{rgb}{0.94, 0.19, 0.22}
\definecolor{radicalred}{rgb}{1.0, 0.21, 0.37}
\colorlet{RED}{red}

\hypersetup{linktoc=all,
	colorlinks, linkcolor={palatinateblue},
	citecolor={brightpink}, urlcolor={amaranth}
}
\newcommand{\changeurlcolor}[1]{\hypersetup{urlcolor=#1}}

\renewcommand{\d}[1]{\ensuremath{\operatorname{d}\!{#1}}}

\renewcommand{\d}[1]{\ensuremath{\operatorname{d}\!{#1}}}
\makeatletter
\def\@fnsymbol#1{\ensuremath{\ifcase#1\or $\textleaf$ \or $\PHplaneTree$
		\else\@ctrerr\fi}}%

\makeatother

\def\sideremark#1{\ifvmode\leavevmode\fi\vadjust{\vbox to0pt{\vss
			\hbox to 0pt{\hskip\hsize\hskip1em
				\vbox{\hsize1.5cm\tiny\raggedright\pretolerance10000
					\noindent #1\hfill}\hss}\vbox to8pt{\vfil}\vss}}}

\begin{document}
	\title{How Not to Extract Information from Black Holes: \\ Cosmic Censorship as a Guiding Principle}
	
	\author{Sofia Di Gennaro}
	\email{sofia.digennarox@gmail.com}
	\affiliation{Center for Gravitation and Cosmology, College of Physical Science and Technology, Yangzhou University, \\180 Siwangting Road, Yangzhou City, Jiangsu Province  225002, China}

	\author{Yen Chin \surname{Ong}}
	\email{ycong@yzu.edu.cn}
	\affiliation{Center for Gravitation and Cosmology, College of Physical Science and Technology, Yangzhou University, \\180 Siwangting Road, Yangzhou City, Jiangsu Province  225002, China}
	\affiliation{Shanghai Frontier Science Center for Gravitational Wave Detection, School of Aeronautics and Astronautics, Shanghai Jiao Tong University, Shanghai 200240, China}

	\begin{abstract}
		Black holes in general relativity are commonly believed to evolve towards a Schwarzschild state as they gradually lose angular momentum and electrical charge under Hawking evaporation. However, when Kim and Wen applied quantum information theory to Hawking evaporation and argued that Hawking particles with maximum mutual information could dominate the emission process, they found that charged black holes tend towards extremality. In view of some evidence pointing towards extremal black holes being effectively singular, this would violate the cosmic censorship conjecture. Nevertheless, since the Kim-Wen model is too simplistic (e.g. it assumes a continuous spectrum of particles with arbitrary charge-to-mass ratio), one might hope that a more realistic model could avoid this problem. In this work, we show that having only a finite species of charged particles would actually worsen the situation, with some end states becoming a naked singularity. With this model as an example, we emphasize the need to study whether charged black holes can violate cosmic censorship under a given model of Hawking evaporation.
		\begin{center}
			
		\end{center}
	\end{abstract}
	
	\maketitle
	
	\section{Hawking Evaporation and Cosmic Censorship}\label{1}
	
	The usual picture of Hawking evaporation of an asymptotically flat\footnote{All black holes discussed in this work are asymptotically flat, and so we will not explicitly mention this hereinafter.} Schwarzschild black hole is that they are governed by the Stefan-Boltzmann law\footnote{In this work we will work in the units $G=c=\hbar=k_B=4\pi\epsilon_0=1$, where $\hbar, G,$ and $c$ are, respectively, the reduced Planck's constant, the Newton's gravitational constant, and the speed of light in vacuum; while $k_B$ denotes the Boltzmann's constant and $\epsilon_0$ the vacuum permittivity.}:
	\begin{equation}\label{SB}
		\frac{\d M}{\d t}= -\alpha a \sigma T^4,
	\end{equation}
	where $M=M(t)$ is the mass of the black hole, and $T=(8\pi M)^{-1}$ is its Hawking temperature.
	Here $a=\pi^2/15$ is the radiation constant, and $\sigma=27\pi M^2$ is the effective area whose radius corresponds to the impact parameter of the photon orbit at $r=3M$ in the geometric optics approximation. 
	Due to scattering at long wavelengths, the actual effective emission surface is smaller. This effect is governed by the so-called ``greybody factor'', denoted here by $\alpha$.
	Depending on the particle species, the numerical value of $\alpha$ is different. Since we are only interested in the qualitative picture, $\alpha$ can be set to unity for simplicity. 
	(For more detailed physics, see the classic work of Page \cite{10.1103}.)
	Solving Eq.(\ref{SB}), one immediately obtains the standard result that a Schwarzschild black hole of initial mass $M_0$ will completely evaporate in a finite time proportional to $M_0^3$. 
	
	Of course, this simple model is likely to fail at late time as spacetime curvature becomes large enough and new physics (quantum gravity) could enter to modify the picture substantially. 
	At the very least, one should expect the emission of new particles beyond the Standard Model.
	Lacking a complete theory of quantum gravity, it is hard to be confident of the final picture. Nevertheless, in the recent years, various studies have converged to suggest that the underlying properties of black holes do change at late time as they undergo Hawking evaporation. In other words, an old black hole behaves rather differently from a young one, a transition that happens long before they reached their final moments. Arguably the most well-known difference is that the Hawking radiation from a young black hole is generally believed to contain no information, but after the black hole has radiated away roughly half of its initial mass, one can start to, in principle, recover the information trapped in the black hole by decoding the quantum information contained in the Hawking radiation (in highly scrambled form); see, e.g. \cite{1409.1231}. This transition between a young black hole and an old one is marked by the Page time \cite{9306083,1301.4995}. For an old black hole, new information that falls into the black hole is quickly scrambled an re-emitted in the Hawking radiation in a much shorter time scale known as the ``scrambling time'' \cite{0708.4025, 0808.2096, 1111.6580, 1307.3458, 1503.01409,1405.7365,1512.04993,1710.03363}. For a Schwarzschild black hole, this is of the order $M \log M$ instead of $M^3$.

	The evaporation of black holes with electrical charge and/or angular momentum is of course more complicated. Nevertheless it is generally expected that black holes would lose angular momentum and electrical charge during the course of Hawking evaporation; if so, \emph{all} black holes would tend to a Schwarzschild state (what happens at the final moments is of course a subject of our ignorance, as previously mentioned). 
	As we will discuss in details in Sec.(\ref{2}), the process of losing charge and angular momentum can be rather nontrivial. The charge-to-mass ratio, $Q/M$, of a charged black hole, for example, can increase towards extremality\footnote{That is, a charged black hole on the verge of becoming a naked singularity -- for Reissner-Nordstr\"om black hole in general relativity, it satisfies $(Q/M)_{\text{max}}=1$.}
	during the course of Hawking evaporation. While this process might take an infinite amount of time (in accordance to the third law of black hole thermodynamics), the worry is that when black holes become near-extremal, a perturbation might bring $Q/M$ to exceed unity and thus destroy the horizon of the black hole, which will expose the singularity within. This would violate the cosmic censorship conjecture \cite{penrose1999,1,1-2,jose} and renders general relativity unpredictable\footnote{More specifically, we mean the ``weak" cosmic censorship conjecture, which essentially states that timelike singularities cannot be naked (i.e., they must be contained inside a black hole horizon), for otherwise it will affect spacetime regions in its causal future.}. Whether such a perturbation exists is, of course, an issue that is still under debate (see the recent review \cite{2005.07032} for some discussions). 
	
	However, problems can arise already when charged black holes become \emph{near}-extremal, even if the horizon is never destroyed to expose the singularity. The reason is as follows: while in classical general relativity charged black holes are described by the Reissner-Nordstr\"om solution, there is evidence that the \emph{interior} of the near-extremal black holes is nothing like what is described by the classical geometry. Instead, as $Q/M$ increases, the singularity becomes closer to the horizon, rendering it eventually ``effectively singular" \cite{1005.2999}. One way to appreciate this picture is completely classical: recall that the inner (Cauchy) horizon is unstable due to an infinite blueshift. This so-called ``mass inflation" is expected to destroy the inner horizon, which might turn it into a new spacelike singularity \cite{9902008}. This phenomenon is expected to persist even at the quantum level \cite{kaminaga} (see, however, \cite{1503.01888} for an opposing argument).
	
	In \cite{1208.3445}, Susskind argued that the growing quantum entanglement of a Schwarzschild black hole with its Hawking radiation causes the singularity to ``migrate" towards the horizon, and eventually intersect with it at sufficiently late time. If correct, then in the Reissner-Nordstr\"om case, even if the inner horizon somehow survives mass inflation \cite{1309.0224}, the singularity can merge with the inner horizon at late time and so for near-extremal black holes we would have essentially the same situation as discussed before. Yet another piece of evidence comes from string theory -- the 4-dimensional low-energy effective theory obtained from heterotic string theory gives the so-called charged dilaton ``GHS" (Garfinkle-Horowitz-Strominger) black hole \cite{ghs,g0,gm}, which can be regarded as a string theoretic correction to the classical Reissner-Nordstr\"om solution. Indeed, an independent argument from area quantization suggests that near-extremal Reissner-Norstr\"om black holes are highly quantum objects \cite{0102061}. Unlike the extremal Reissner-Nordstr\"om  black hole whose horizon remains smooth (if we take the classical geometry prima facie) in the extremal case, the extremal charged GHS black hole is a null singularity, along which spacetime curvature diverges. 
	Thus, both classical and quantum gravity considerations suggest that near extremal charged black holes have arbitrarily large curvature near -- even \emph{outside} -- their horizon. This is as bad as a truly naked singularity, because general relativity cannot describe physics in the region of arbitrarily large curvature. Any spacetime point whose causal past intersects with such region is therefore unpredictable from the (semi-)classical theory.

	In view of this, if the cosmic censorship conjecture is correct, then one should expect that \emph{as a black hole undergoes Hawking evaporation, its parameters should evolve in such a way that it avoids becoming extremal, not just avoid becoming a truly naked singularity}. It is therefore not surprising that we \emph{do} see such a behavior when Hiscock and Weems \cite{HW} modeled the evolution of Reissner-Nordstr\"om black holes under Hawking evaporation taking into account two different regimes: one that favors charge loss and one that favors mass loss. This gives rise to an attractor solution that flows towards the Schwarzschild end state (we will review the details in the next section). However, this is not necessarily the case when a different model of Hawking evaporation is used, such as the one proposed by Kim and Wen \cite{1311.1656}. As per our discussion above, this is likely to be problematic. In this work we will first further analyze the Kim-Wen model in details, pointing out some interesting properties and also the subtle differences compared to the Hiscock-Weems model. In particular we shall point out why the original Kim-Wen model is not realistic. We will then try to improve the Kim-Wen model to see if it can be saved (i.e. to prevent black holes from tending towards extremality), but then conclude that a more realistic model is in fact -- \emph{worse} --  (generic) end states become either extremal or super-extremal (i.e., a naked singularity), or a highly charged remnant \cite{Chen}. We will end with some discussions about the role of cosmic censorship in semi-classical and quantum gravity in general; but more specifically, on the role of cosmic censorship as a guiding principle to let us rule out models of Hawking evaporation beyond Eq.(\ref{SB}).

	\section{How Charged Black Holes Evaporate}\label{2}
	
	\subsection{Hiscock-Weems Model: Extremality Is Never Reached}
	A Kerr black hole is characterized by its mass $M$ and angular momentum $J$. It is often more convenient to work with the angular momentum parameter $a=J/M$. 
	The evolution of Kerr black holes under Hawking evaporation is straightforward -- they spin down. It is true that if a Kerr black hole only emits scalar particles, it will evolve towards $a/M \approx 0.555$, however once higher spin particles are added, the final state would be $a/M \to 0$ \cite{9710013, 9801044}; see also \cite{115295}. As such, we do not consider angular momentum in this work but instead focus on the effect of electrical charge.
	
	Depending on the charge-to-mass ratio $Q/M$ (we assume $Q>0$ without loss of generality) of a Reissner-Nordstr\"om black hole\footnote{Here we only consider isolated black holes. In an actual astrophysical environment discharge would be more efficient due to infalling matter, but even then small charges can still give rise to non-negligible effects \cite{1904.04654}. In the context of black hole evaporation, small charge-to-mass ratio can grow, so for a sufficiently isolated black hole we cannot ignore the charges. 
	}, discharge can be quite efficient, or it can be very slow \cite{HW, gibbons}. For sufficiently large black holes $M \gg Q_0 := q_e/(\pi m_e^2)$, where $q_e$ and $m_e$ denote the charge and mass of an electron\footnote{Since we assumed $Q>0$, it is actually the positron that is preferentially emitted (like charges repel).}, Hiscock and Weems argued that the evaporation can be modeled by the coupled ordinary differential equation
	\begin{equation}\label{dMdt}
		\frac{\d M}{\d t} = -a \alpha \sigma T^4 + \frac{Q}{r_+}\frac{\d Q}{\d t},
	\end{equation}
	where the charge loss rate
	\begin{equation}\label{dQdt}
		\frac{\d Q}{\d t} \approx -\frac{e^4}{2\pi^3 m_e^2}\frac{Q^3}{r_+^3}\exp\left(-\frac{r_+^2}{Q_0Q}\right)
	\end{equation}
	is obtained from an approximation of the Schwinger formula in quantum electrodynamics \cite{HW, Schwinger, g}, which governs charged particle production in the presence of a strong electric field. 
	(See also \cite{Khriplovich,1202.3224,2003.07016}.)
	The threshold beyond which pair production is effective is given by the Schwinger critical field $E_c:={m_e^2 c^3}/{q_e \hbar} = 1.312 \times 10^{18}~\text{V/m}$ in SI units. In the regime of validity of the model, the Schwinger effect is actually suppressed: the electric field never exceeds the critical value as long as the black hole never exceeds its extremal charge.
	In this model since the mass of the black hole is large enough, its Hawking temperature, 
	\begin{equation}
		T=\frac{\sqrt{M^2-Q^2}}{2\pi (M+\sqrt{M^2-Q^2})^2},
	\end{equation}
	is sufficiently low not to emit heavier charged particles. Massless particle emission is governed by the Stefan-Boltzmann term. 
	
	Hiscock and Weems found that if the initial charge-to-mass ratio of a black hole is sufficiently large, the black hole steadily discharges. However, if $Q/M$ is initially small, then its value will \emph{increase at first}, since it is losing much more mass than charge. It is possible that $Q/M$ comes very close to unity, i.e., the black hole becomes near extremal. Nevertheless, $Q/M$ cannot keep on increasing indefinitely. On the contrary, $Q/M$ will decrease once discharge becomes more efficient \cite{1909.09981}, when $\d M/\d t \sim \d Q/\d t$. This gives rise to the attractor behavior in Fig.(2) of \cite{HW}, part of which is re-produced in Fig.(\ref{Hiscock-Weems}) below. 
	We remark that the charge loss term, \emph{despite being suppressed}, is crucial -- otherwise the emission of only chargeless particle could lead to a violation of cosmic censorship as well, as shown recently by Hod \cite{2102.05519}. Although the emission of charged particles is suppressed in this regime, its inclusion in the Hiscock-Weems model makes all the difference.  
	Finally, we remark that although the curves get very close to the attractor, they never actually cross, since the evolution under differential equations is of course unique.
	
	\begin{figure}[h!]
		\centering
		\includegraphics[scale=0.55]{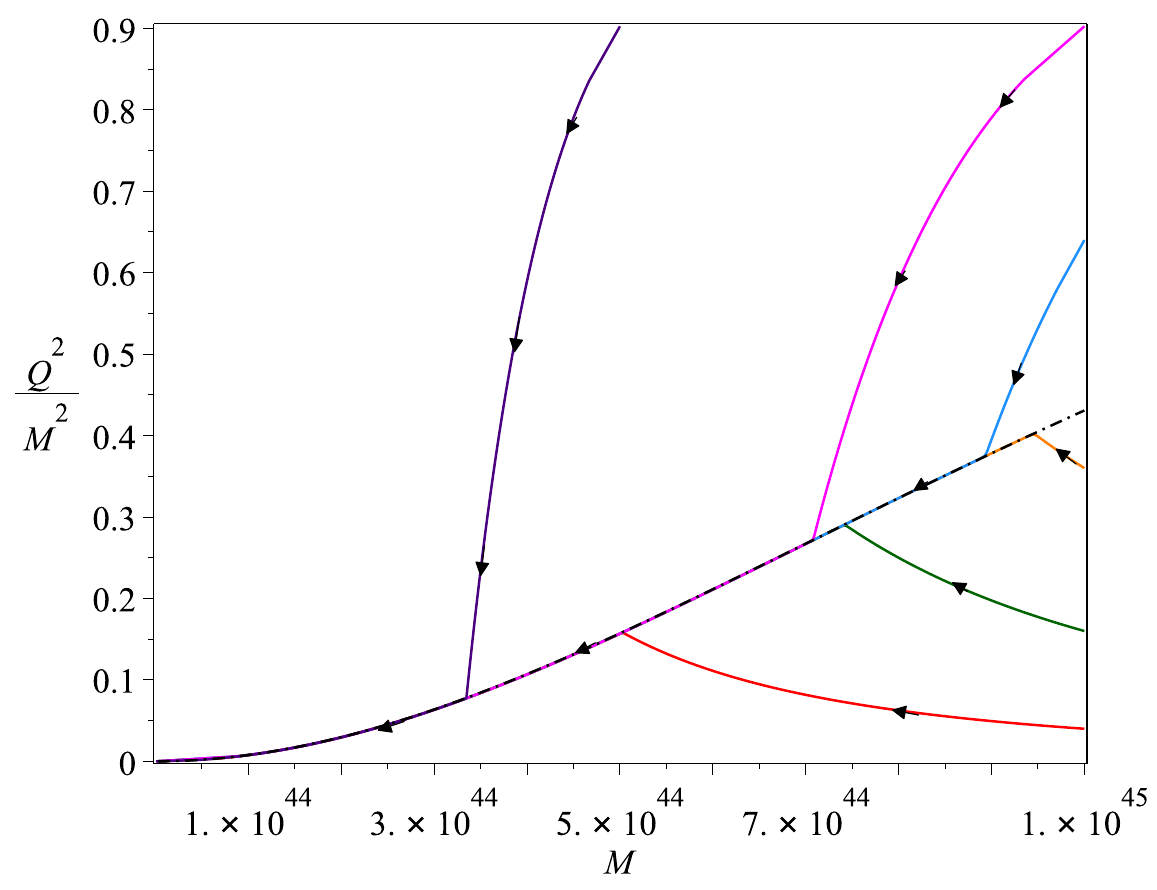}
		\caption{Evolution of Reissner-Nordstr\"om black hole under Hawking evaporation in the Hiscock-Weems model. Given any initial condition $M_0$ and $Q_0$, we can track how the ratio $Q/M$ evolves in this plot.
			The attractor (dash-dot black curve) is characterized by \cite{1909.09981} $\d M/\d t \sim \d Q/\d t$, it tends to $(Q/M)^2=1$ when $M \to \infty$. This means that sufficiently large black holes can approach but cannot reach extremality; trajectories always eventually turn away from extremality. In the regime of validity of the model ($M \gg Q_0 $), the black hole always -- eventually -- evolves towards a Schwarzschild state.}
		\label{Hiscock-Weems}
	\end{figure}
	
	The Hiscock-Weems model \emph{suggests} that the end state is Schwarzschild, which by our discussion in Sec.(\ref{1}), will completely evaporate away in a finite time.  
	This is only suggestive because we can no longer trust the model once $M \lesssim Q_0$ (which is about $1.7 \times 10^5$ solar masses \cite{HW}, that is to say, the model is only good for supermassive black holes). Nevertheless, if no new physics enters the picture, there seems to be no reason to suspect that $Q/M$ will rise again for small $M$. This should be the case until $M$ becomes extremely small at the very late stage of the evaporation and the temperature scale is high enough for (potentially) new physics to kick in.
	
	Note that as emphasized in \cite{1907.07490,1909.09977,1909.09981}\footnote{We remark that the value of $\hbar$ in these papers -- though written correctly -- was off by a factor of 50 or so in the numerical codes due to a typo. This does not affect the main conclusions.}, the fact that charged black holes never become extremal in this model is not only consistent with the third law of black hole thermodynamics, it is also an evidence for the validity of the cosmic censorship -- if a black hole becomes extremal, a perturbation may destroy the horizon and renders the singularity naked (though whether such perturbations exist remains unclear). The role of the third law in upholding the cosmic censorship was already pointed out by Davies in 1977 \cite{davies}. However, note that the attractor behavior is crucial; the third law itself is not enough: for a charged dilaton black hole the extremal black hole is itself a null singularity -- curvature (Kretschmann invariant) gets arbitrarily large near extremality, which effectively already violates the spirit of cosmic censorship. As per our discussion in Sec.(\ref{1}), this could also be the case of a more realistic extremal Reissner-Nordstr\"om black hole. The third law can only guarantee that a black hole takes an infinite amount of time to reach extremality, but does not forbid it to become arbitrarily close in a finite (albeit extremely long) time. 
	Then, when it was shown that a modification to the charged particle emission rate in the presence of dilaton field, as required by quantum field theory, is precisely the ``minimal requirement" needed to uphold
	cosmic censorship in the Hiscock-Weems model, it was a strong evidence for the conjecture's validity \emph{even in semi-classical gravity} \cite{1907.07490}. We will come back to this interesting issue in the Discussion section.  
	
	Let us now review the Kim-Wen model of charged black hole evaporation, which behaves in a completely different manner and potentially violates cosmic censorship (we will argue that it does).
	
	\subsection{Kim-Wen Model: The Role of Quantum Information}{\label{sub2}}
	
	Kim and Wen argued for a completely different picture: that the end state is an extremal black hole \cite{1311.1656}.
	Kim and Wen actually considered Hawking radiation as quantum tunneling \emph{\`a la} Parikh-Wilczek \cite{9907001,1609317,699377}, with the additional criteria that the emission is dominated by those with maximum mutual information (MMI). 
	In more details, the Parikh-Wilczek tunneling picture considered Hawking radiation as the result of particles tunneling out from the black hole, whilst enforcing the conservation of energy\footnote{This is not a trivial statement. In general relativity energy (of the matter sector) is strictly conserved only when there is a timelike Killing vector, or equivalently in the language of field theory (Noether's theorem), when there is a time translation symmetry -- which clearly is not the case if a black hole is evaporating. Here we follow the usual assumption -- which is \emph{justified a posteriori} from the sensible results obtained via the Parikh-Wilczek formalism throughout the literature -- but feel the need to point out this subtlety.}, that is, by assuming the black hole mass $M$ is the total energy of all the Hawking particle emitted. The radiation is not exactly thermal in this picture as two consecutive emissions are not independent, as we are going to see in the next paragraph.\\ 
	
	The tunneling probability is given by $\Gamma = \ln S$, where $S$ indicates the mutual information entropy, given by the standard definition:
	\begin{equation}
		S(A:B):=S(A)+S(B)-S(A,B) \label{mutual info}
	\end{equation}
	for two quantum subsystems $A$ and $B$.
	Then, for two consecutive emissions it corresponds to (see \cite{0903.0893}): 
	\begin{equation}
		S_{\text{MI}}(X;x_1,x_2):=S(x_2|x_1)-S(X;x_2), \label{mutual info 2}
	\end{equation}
	where $X$ represents the properties (``hairs'') of the total black hole (for instance the mass, charge and angular momentum), while $x_1$ and $x_2$ are those of the emitted particles. The notation $S(x_2|x_1) := \ln \Gamma(x_2|x_1)$ refers to the conditional probability of emission, which is the probability of emitting $x_2$ given the previous emission of $x_1$. To see this more explicitly, let us apply it to the Reissner-Nordstr\"om black hole that is of interest for this work. This yields
	\begin{align}
		S(M,Q;m,q) & :=\pi \left\{ \left[M-m + \sqrt{(M-m)^2-(Q-q)^2}\right]^2 -\left(M+\sqrt{M^2-Q^2}\right)^2\right\} \nonumber\\
		& \equiv \pi \left( r_{(1)}^2 -r_{(0)}^2 \right), \label{rs mutual info}
	\end{align}
	where the notation $r_{(n)}$ represents the horizon radius after the emission of $n$ particles, which implies that the entropy is a function of $M-nm$ and $Q-nq$ instead of $M,Q$. Then, for the total mutual information of Eq.\eqref{mutual info 2} we have:
	\begin{equation*}
		S_{\text{MI}}(M,Q;m_1,q_1,m_2,q_2):=S(M-m_1, Q-q_1;m_2,q_2)-S(M,Q;m_2,q_2).
	\end{equation*}
	If the two particles are of the same kind, we can use $m_1 = m_2 = m$ and $q_1 = q_2 = q$ to write:
	\begin{equation}
		S_{\text{MI}} := \pi\left( r_{(2)}^2 -2r_{(1)}^2 + r_{(0)}^2 \right). \label{entropy 2}
	\end{equation}	
	Moreover, it can be proven, using Jensen's inequality, that the quantity in Eq.\eqref{mutual info} is necessarily non-negative. For completeness, let us plot the mutual information given by Eq.\eqref{rs mutual info} (this was already done in \cite{1311.1656}) to see the bounds its non-negativity imposes on the ratios $Q/M$ and $q/m$, assuming  the emission of
	identical particles with mass $m$ and charge $q$.
	\begin{figure}[h!]
		\centering
		\begin{subfigure}
			\centering
			\includegraphics[scale=0.70]{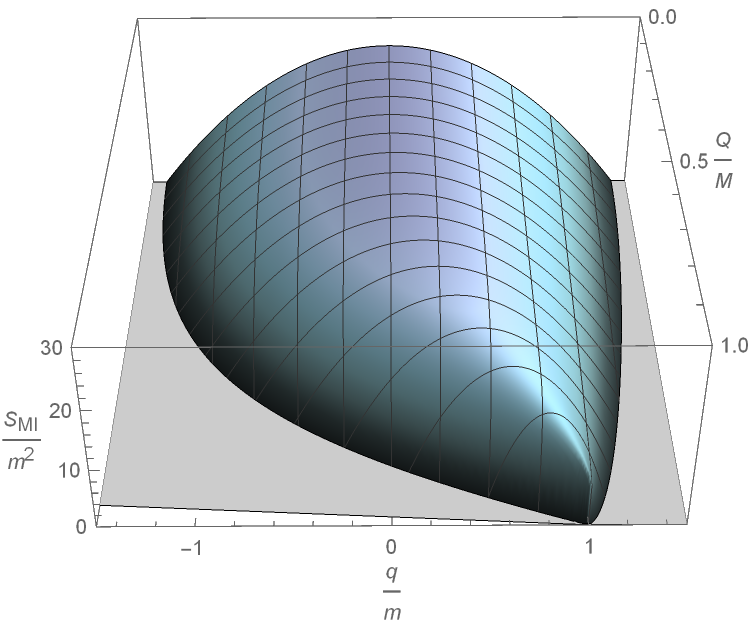}
		\end{subfigure}
		\begin{subfigure}
			\centering
			\includegraphics[scale=0.80]{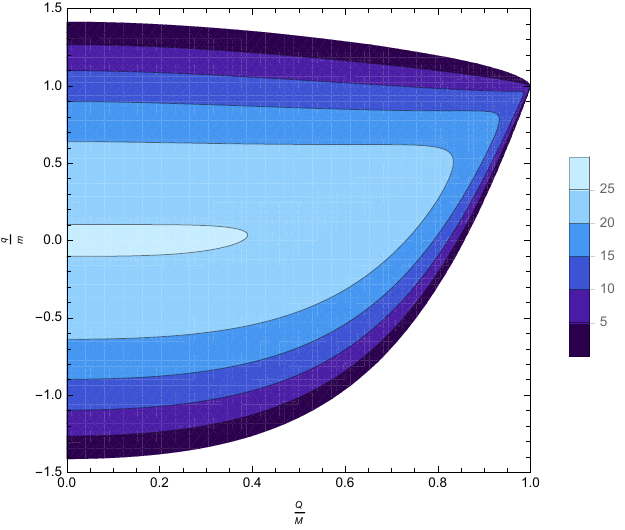}
			\caption{{Mutual information in the emission from a Reissner-Nordstr\"om  black hole with $M/m = 10^2$. The plots consider only the emission of identical particles with mass $m$ and charge $q$. The mutual information is presented as a function of both $Q/M$ and $q/m$. \textbf{Left:} The 3-dimensional plot; \textbf{Right:} The corresponding contour plot.}}
			\label{fig:mutual info 3d}
		\end{subfigure}
	\end{figure}
	
	In Fig.(\ref{fig:mutual info 3d}), following \cite{1311.1656}, we included negative $q/m$ ratios to better display the behavior and the maximum  of the mutual information function. It must be noted, though, that we have assumed $Q>0$, so only positive values for $q$ are allowed. Indeed, when a charged pair is produced in the electrical field of the black hole, the particle with the same charge as the black hole is repulsed to infinity, while the other one is absorbed, thus reducing the black hole's charge. Therefore, for $Q>0$ we only consider the emission and tunneling of positive electric charges. This is the same assumption made in the Hiscock-Weems model. 
	
	From the contour plot we can observe the bounds on $q/m$ and $Q/M$ better: we notice that all values of $Q/M \in \left( 0,1 \right) $ are allowed only for $q/m=1$. As $Q/M$ tends towards $0$, it is around $\sqrt{2}$ (in the large mass limit, as shown in \cite{1311.1656} this upper bound is saturated). When $q/m \approx 0$, higher $Q/M$ values are not possible anymore. The lowest $Q/M$ is about $0.86$ for $q/m=0$. This value is intriguing, as we will comment on later.

	The explicit entanglement between the Hawking particles in the Schwarzschild case allows the mutual information to exhibit the behavior of the Page curve \cite{9306083,1301.4995}, i.e., $S_{\text{MI}}$ first grows then decreases at about the midpoint of the evaporation process \cite{1311.1656}. See also \cite{1308.2386}. Whether this can completely resolve the information paradox remains controversial \cite{1108.0302,1210.2048,1908.09669}, but is of little relevance to our present work.

	Kim and Wen showed that if one further imposes the MMI principle,
	the evaporation happens through the progressive emission of particles with the following charge-to-mass ratio \cite{1311.1656}:
	\begin{equation}
		\gamma := \left(\frac{q}{m}\right)_{\text{optimal}}= \frac{Q^3}{M^3+(M^2-Q^2)^{\frac{3}{2}}}. \label{gamma}
	\end{equation}
	This result comes from calculating the $q/m$ ratio that maximizes the mutual information entropy of the emitted particles, assuming $M, Q \gg m,q$.
	While the MMI principle\footnote{This is better known as the ``infomax principle'' in the parlance of information theory and neural networks \cite{linsker}. In the zero-noise limit it is equivalent to the principle of redundancy reduction in biological sensory processing \cite{barlow, redlich, parga}. For a review of the applications of information theory in evolutionary biology, see \cite{1112.3867}, in which the notion of entropy and information of molecules and proteins are discussed.} is an assumption, it is nevertheless a well-motivated one. It is essentially the statement that black hole evaporation is not a random process, but an \emph{optimized} one that allows information to escape as fast as possible. This is in line with the evidence found thus far that suggests black holes are the fastest computers in Nature \cite{0708.4025, 0808.2096, 1111.6580, 1307.3458, 1503.01409,1405.7365,1512.04993,1710.03363}. The question is whether such a model is realistic. We first note that under MMI, particle emission is governed by mutual information and not the temperature, thus it is entirely possible that emission of massive charged particle is favored over a massless neutral one, which is a huge deviation from the thermal emission of Hawking radiation. Usually we expect corrections to Hawking's results to be small, if any, and thus on this ground alone the model is already somewhat uncomfortable. Nevertheless, putting this prejudice aside\footnote{The position one may take varies. For example, assuming the usual AMPS assumptions lead to the ``firewall'', which is also a huge deviation from what one expects from QFT on curved spacetime; an alternative way to evade firewall by a different mechanism to extract quantum information may therefore deviate from the usual thermal one by an appreciable amount, at least in certain regime. We are playing the devil's advocate here and are not seriously advocating for this scenario, of course. In any case, the point of this work is only to use the Kim-Wen model as an explicit example to illustrate the idea that cosmic censorship plays a crucial rule in checking the feasibility of any proposed model of Hawking radiation, some of which may in principle have smaller deviation from the standard form but nevertheless affect how $Q/M$ evolves in certain regimes of the parameters.}, let us see what are the ramifications of the Kim-Wen model. 
	
	We will first illustrate the Kim-Wen model with a simple example. Let $n \in \mathbbm{N}$.
	Let us consider a black hole of mass $M=nm$, which emits a particle of mass $m$ at each step. This results in the following change in the charge-to-mass ratio for the black hole:
	\begin{equation}
		\frac{Q}{M}\quad\longrightarrow\quad \frac{\frac{Q}{m}-\gamma}{\frac{M}{m}-1}. \label{eq1}
	\end{equation}
	We can then, with loss of generality, assume $m=1$ for all the particles and plot the evolution, considering that every time a particle is emitted, in Eq.\eqref{gamma}, $Q$ and $M$ are updated with the new values. The result for $n=100$ is plotted in Fig.(\ref{Kim Wen})\footnote{In all the subsequent plots of this paper, the blue curve represents the curve with initial charge $Q_i=0.8$ $M_i$, the gold, green and red curves correspond to $Q_i = 0.6,0.4,0.2$ $M_i$, respectively.}, which reproduces the behavior found in Fig.(4) of \cite{1311.1656} (including the effect of angular momentum does not change this final fate \cite{1403.3986}):
	
	\begin{figure}[h!]
		\centering
		\includegraphics[scale=0.90]{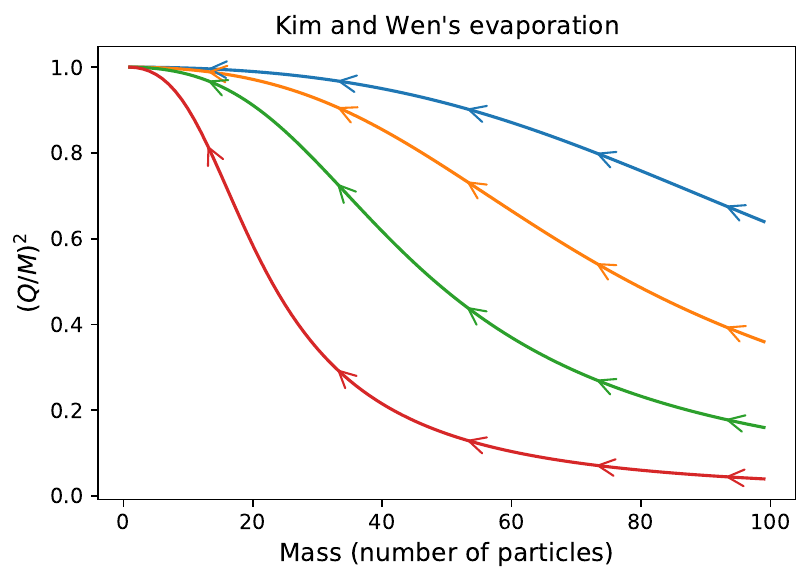}
		\caption{Evolution of Reissner-Nordstr\"om black hole under Hawking evaporation in the Kim-Wen model: all black holes eventually tend to extremality.}
		\label{Kim Wen}
	\end{figure}
	
	Why is there such a stark discrepancy in the results between Hiscock-Weems and Kim-Wen models? The answer is of course: because their underlying assumptions are different. Specifically the MMI principle is absent in the Hiscock-Weems model. The main difference here is that the MMI principle (together with the requirement that the mutual information is non-negative) picks out what type of particle is to be emitted at a certain step, whereas the Hiscock-Weems model only considers emission of massless particles and electron/positron. In fact, since charged Standard Model particles all have $q/m\gg 1$, they \emph{cannot} be emitted under the Kim-Wen model. This is because non-negativity of the mutual information imposes an upper bound on the $q/m$ ratio of emitted particle to be $q/m=\sqrt{2}$ \cite{1311.1656}. A lower bound appears when $Q/M \gtrsim 0.86$ \cite{1311.1656}. We note that curiously this is very close to the value $Q/M=\sqrt{3}/2\approx 0.866$ (``Davies point''), beyond which the specific heat $C:=\d M/\d T$ for a Reissner-Nordstr\"om black hole with a \emph{fixed} charge becomes positive \cite{davies}, despite the fact that the specific heat under the Kim-Wen model is always negative \cite{1311.1656} (See more details in Appendix \ref{appendixa}). We do not know whether there is a deeper meaning to this coincidence.

	It is of course possible that both models are correct, but they are applicable in different regimes. We have already explained that the Hiscock-Weems model is good for sufficiently large astrophysical size black holes but does not apply for small black holes. Our example above definitely counts as a small black hole since $n$ is only 100. Unfortunately numerical limitation does not allow us to check whether the Kim-Wen model reproduces the same behavior as in Hiscock-Weems {(our results hold even when we increased $n$ to $10^6$)}, since that requires $n \sim 10^{60}$ (ratio of a solar mass to electron mass, as a quick approximation) or more. 
	
	It is likely that the Kim-Wen model would not give the same behavior as that of Hiscock-Weems model for large black holes, for which we would expect that the standard picture of Hawking radiation as blackbody holds (any deviation from the Planck spectrum would be small -- note that even the Hiscock-Weems model is not exactly thermal due to the Schwinger term), namely temperature governs the species of Hawking particles that can be emitted. As the black hole becomes very small, its quantum nature would become more dominant and maybe only then the MMI principle will take over. {(Note, however, in neither of these models can we be absolutely confident about the very final moment when the black hole is extremely small. For the Kim-Wen model this is because $\gamma$ in Eq.(\ref{gamma}) no longer holds in the Planckian regime.)} Such an optimization to release information as soon as possible is in line with the notion that after the Page time, quantum information that falls into the black hole is quickly re-emitted in a timescale set by $T^{-1} \log S$, the ``scrambling time'' \cite{0708.4025, 0808.2096, 1111.6580, 1307.3458, 1503.01409,1405.7365,1512.04993,1710.03363}, where $S$ denotes the Bekenstein-Hawking entropy of the black hole (a quarter of the horizon area). This is also consistent with recent findings that terms beyond leading order in the Hawking temperature can have pronounced effects at late time \cite{1906.01735}. However, we can in principle extract quantum information in many ways, which are all in line with the aforementioned general results (that require the recovery of quantum information in some efficient way after the Page time). Indeed, if we impose some principle $P$ to govern at each step what sort of particles should be emitted from a black hole, it would of course affect how the charge-to-mass ratio evolves. The fact that the Kim-Wen model led to an extremal end state suggests that it should be ruled out as a viable information extraction principle. However, in order to reach such a conclusion we must show the conclusion is robust: we will first try to ``save'' the model, and in doing so show that the improved version is also problematic (actually worse).

	\section{An Improved Kim-Wen Model}\label{3}
	\subsection{A Discrete Set of Particles}
\begin{figure}[h!]
		\centering
		\includegraphics[scale=0.7]{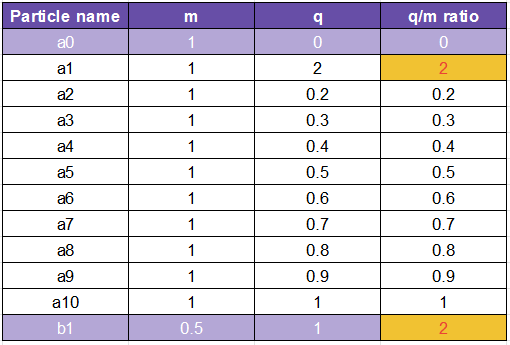}
		\caption{Table of all the available (fictitious) particles and their properties. The first column shows the name of the particle, followed by its mass $m$, its charge $q$ and its $q/m$ ratio (shaded in dark yellow if it is above $\sqrt{2}$). The particles shaded in violet are special, for instance $a0$ is the neutral particle and $b1$ is the only one with a different mass, it is included to test the effect of a smaller mass for particles with $q/m > \sqrt{2}$.}
		\label{fig:table1}
	\end{figure}

\begin{figure}
		\centering
		\includegraphics[scale=0.7]{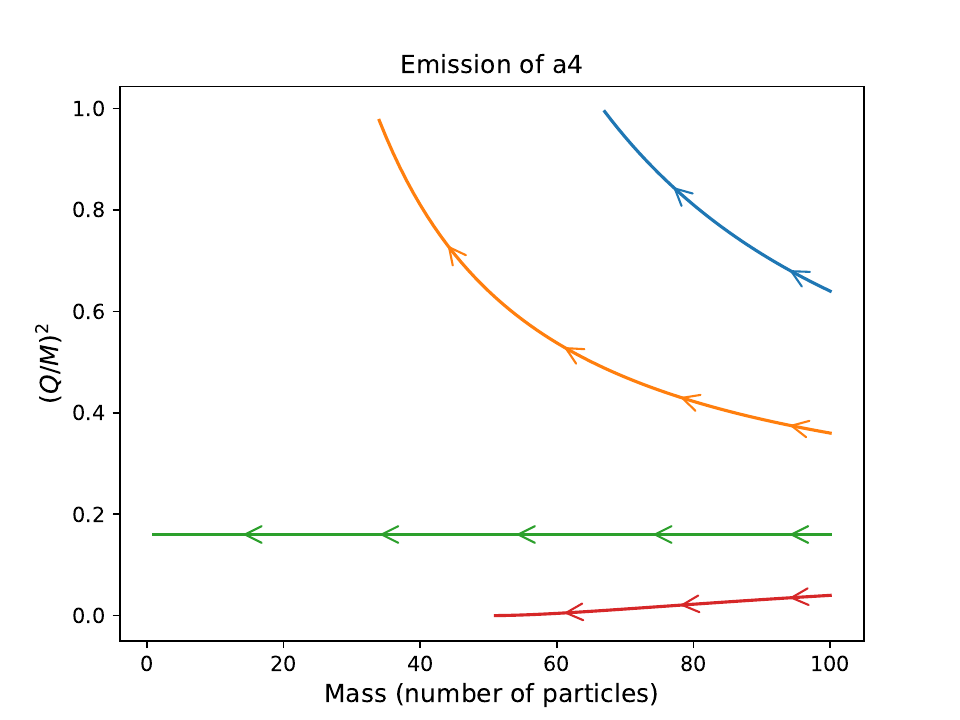}
		\caption{Evolution of the black hole when only the particle $a4$ is present. The  horizontal axis represents the total mass of the black hole, as well as the number of particles (we assume all the particles have $m=1$, see Table (\ref{fig:table1})). The blue curve corresponds to an initial charge-to-mass ratio of the total black hole of $Q_i/M_i = 0.8$. The yellow, green and red curves correspond to $Q_i/M_i = 0.6,0.4,0.2$ respectively. We shall use these colors to denote the curves with these initial values in all the plots hereinafter.}
		\label{fig:em4}
	\end{figure}

	Another major difference between the Kim-Wen model and the Hiscock-Weems model is that the latter only considers massless particles as well as electron/positron (since the emissions of other charged particle are suppressed in the regime of validity of the model), whereas the Kim-Wen model considers a \emph{continuous} set of emitted particles, so that the optimal ratio $\gamma$ from Eq.\eqref{gamma} corresponds exactly to the $q/m$ ratio of the particle. To investigate how the Kim-Wen model behaves in a more realistic (though still hypothetical) universe, we analyze what happens if we discretize the set of available particles. We introduce the fictitious particles $\left\{a0,a1,\cdots ,a{10}, b1\right\}$, each corresponding to a different $q/m$ ratio; we list their specifics in the table given in Fig.(\ref{fig:table1}).
	In particular, we have included a chargeless particle $a0$ that mimics neutrinos in the Standard Model, which can be massless or massive. It can also mimic massless particles like photon and graviton (since massless particles contribute to Hawking evaporation because they carry energy), which in this simple toy model we treat as an effective mass.\\	
	\noindent First we analyze the case in which only one particle at a time is available to be chosen for the evaporation of the black hole. In particular, we start by analyzing what happens when we have only a particle with $q/m>\sqrt{2}$, like $a1$. We confirm the result that we anticipated when discussing Fig.(\ref{fig:mutual info 3d}), that whenever the mutual information is negative that particular emission channel does not occur.
	On the other hand, the other particles $\left\{a1, \ldots, a10\right\}$ all have $q/m \leqslant 1$; their behavior can be best exemplified by the plot in Fig.(\ref{fig:em4}), in which the particle $a4$ is the only one present. 
	In this plot the three possible outcomes are quite clear: the green line (corresponding to $Q_i/M_i = 0.4$, see caption of Fig.(\ref{fig:em4})) stays constant, the curve below it goes to $Q=0$ and the curves above it all rise to extremality.	
	The reason why this happens follows from Eq.\eqref{eq1}: when the particle exactly satisfies $q/m = Q/M$, the total charge-to-mass ratio is left unchanged by the emission of such particle, thus creating a constant line in the evolution plot. Similarly, if $q/m < Q/M$, the latter ratio tends to increase and we see a rising behavior in the plots, and the opposite is true for $q/m > Q/M$. It is now obvious that the particle $a0$, being chargeless, has the lowest $q/m$ possible and thus will inevitably lead to a rising behavior.
	
	The next point that is worth noting is that the rising curves in Fig.(\ref{fig:em4}) (blue and yellow) stop when the black hole reaches the extremal state $Q = M$, as the result of the mutual information becoming negative. In this case, the MMI prevents the ratio $Q/M$ from uncontrolled growth (i.e. from becoming super-extremal; though as we have argued that this is still problematic). After this evaporation process the black hole would become an extremal remnant \cite{Chen} that has no way of evaporating; whereas if it follows the red curve it ends up as a neutral ``Schwarzschild-like remnant'' (at some point the evaporation stops). The green line is somehow in the middle of these two situations, resulting in the black hole fully evaporating and leaving only a single particle $a4$.	
	
	\begin{figure}[h!]
		\centering
		\begin{subfigure}
			\centering
			\includegraphics[scale=0.45]{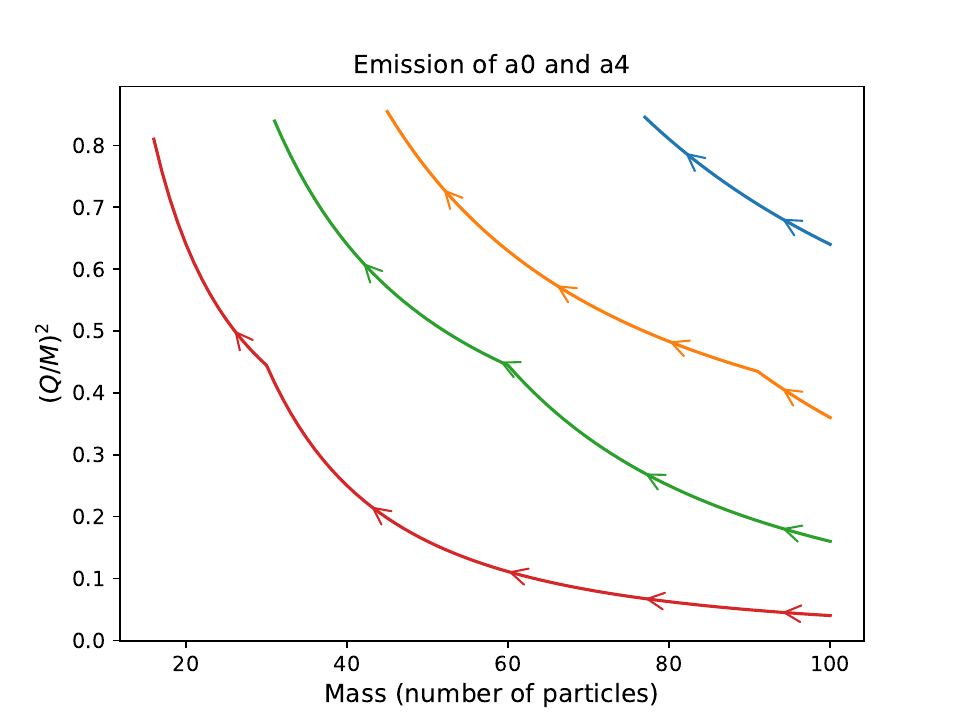}
		\end{subfigure}
		\begin{subfigure}
			\centering
			\includegraphics[scale=0.45]{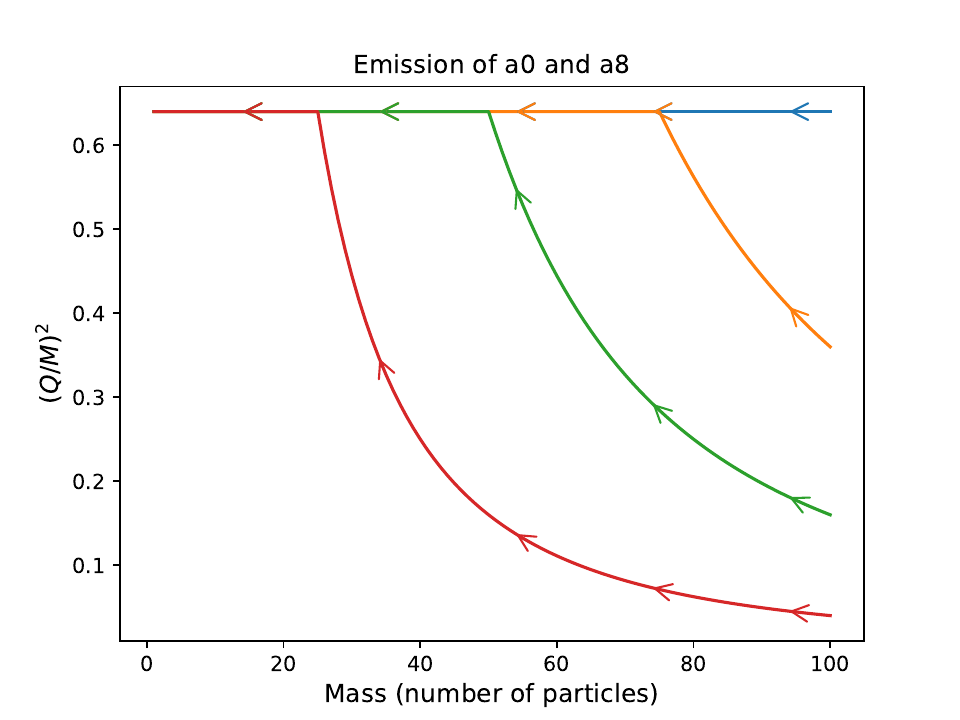}
			\caption{\textbf{Left:} Emission of the particles $a0$ and $a4$. \textbf{Right:} Emission of the particles $a0$ and $a8$.}
			\label{fig:a4a8}
		\end{subfigure}
	\end{figure}
	\begin{figure}[h!]
		\centering
		\begin{subfigure}
			\centering
			\includegraphics[scale=0.45]{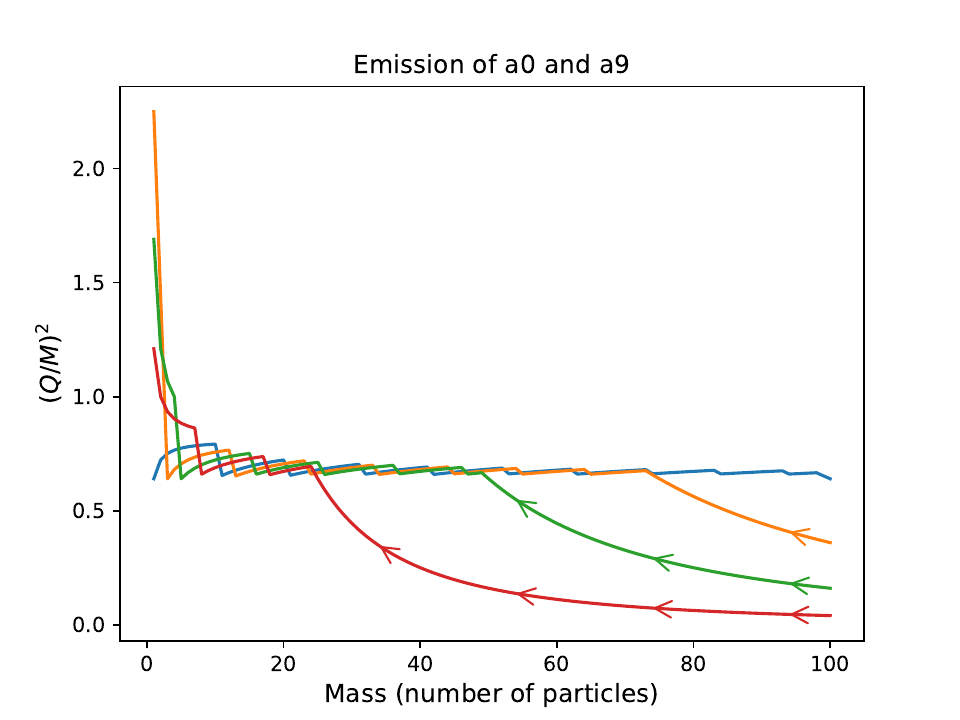}
		\end{subfigure}
		\begin{subfigure}
			\centering
			\includegraphics[scale=0.45]{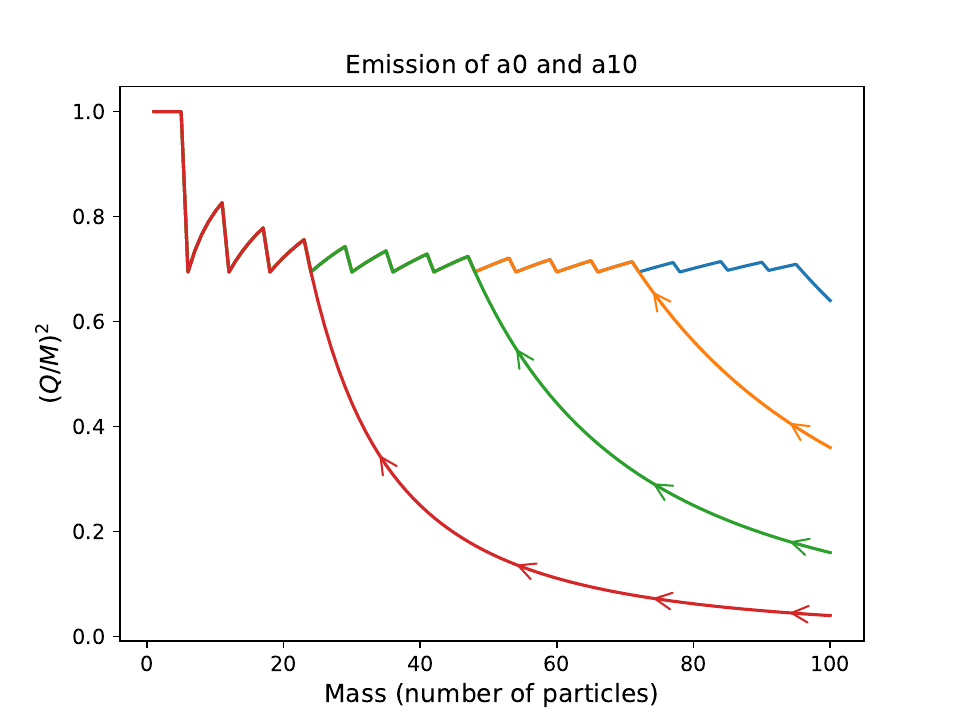}
			\caption{\textbf{Left:} Emission of the particles $a0$ and $a9$. \textbf{Right:} Emission of the particles $a0$ and $a10$.}
			\label{fig:a9a10}
		\end{subfigure}
	\end{figure}
	Of course, none of the cases above is really physical, as in a more realistic universe (hypothetical or not) we expect to have more than one species of particle. In particular, a theory with charged black holes should also at least contain particles such as gravitons and photons, which are chargeless. For this reason, in the next tests of the Kim-Wen model, we always include the neutral particle $a0$ and then add a set of the other particles, covering all the possible combinations. We implement the MMI optimization in the following way: at each step of the evaporation, we calculate the value of $\gamma$ and emit the particle whose $q/m$ ratio is the closest to it. If the particle cannot be emitted because the charge is not sufficient, the next one is chosen instead.
	
	To start, we look at what happens when we have the neutral particle and only one of the charged ones. In the first plot in Fig.(\ref{fig:a4a8}) all the curves rise towards extremality, mainly because of $a0$ (emitted in the first part of the evaporation, followed by $a4$). The curves stop before reaching $Q=M$ because of the requirement that the mutual information be non-negative.
	All the particles up to $a7$ behave in a similar fashion, except for $a1$, which cannot be emitted at all because of the negativity of mutual information (see Fig.(\ref{fig:mutual info 3d})). In other words, these are charged remnants, which are usually thought to be problematic, though  there are different points of view \cite{Chen}, so we may not want to rule out the model based on this result alone.
	
	On the other hand, for $a8$ we get to the case of the second plot of Fig.(\ref{fig:a4a8}), where all the curves become constant; this is explained by the fact that when $a8$ starts to be emitted, the total $Q/M$ ratio is exactly $0.8$. This is clearly a non-generic situation.
	
	Next are the particles $a9$ and $a10$, shown in Fig.(\ref{fig:a9a10}). Starting with $a10$, we notice that the curve is oscillatory when the particles $a0$ and $a10$ are emitted alternatively and it eventually ends at extremality. In the plot of $a9$, we can see a similar oscillation, but the end state is drastically different: three of the curves are able to pass \emph{beyond} extremality even if the non-negativity condition of mutual information is imposed\footnote{The charge-to-mass ratio $\gamma$ is related to the black hole parameters via Eq.(\ref{gamma}), which is not defined for $Q^2>M^2$. However, since the mutual information is still positive it seems reasonable that particle can be emitted, so we let the evolution proceeds by manually setting $\gamma$ to be say, 0 or 1; the values do not affect the results.}. 
	At this point let us remark that the evolution curves are obtained by joining discrete points, thus resulting in the curves crossing each other. However, none of the data point for different curves coincide. In this sense, given an initial condition, the discrete evolution is still unique. 
	
	Thus far we have only performed a relatively simple examination using a specific particle together with the chargeless one, so next we performed a more realistic test, running the algorithm for the evaporation involving all the possible combinations of $\left\{a0, a1, \cdots, a10\right\}$. 
	It turned out that the only ``good plots'' (that neither tend towards extremality nor go beyond it) are the aforementioned set $\left\{a0,a8\right\}$ and the set $\left\{a0, a2, a4, a9\right\}$, which tend to a constant value of the charge-to-mass ratio. This occurs in only two situations among all the possible combinations of particles, which means that it is non-generic.
	
	
	
	Moreover, in all the possible combinations, the particle $a1$ is never chosen, not even as a last resource when no other particles can be emitted. We performed a test using the particle $b1$, which has the same charge-to-mass ratio as $a1$, but its mass is half of it. What we found is that even this particle is not emitted in any circumstance.
	
	The last test consists of using the mutual information entropy to determine whether a particle can be emitted of not. First, the optimal ratio $\gamma$ is calculated; then, if the chosen particle would generate a negative value for the mutual information, the next one with the closest charge-to-mass ratio to $\gamma$ is picked instead. If none of the particles is suitable, then the evaporation stops. Doing this, we observe that particles such as $a1$ and $b1$ \emph{can} be emitted. However, this does not change the overall result we obtained from the previous analysis as no new ``good" plots are discovered. {Lastly, we remark that the qualitative results above also do not change if we include particles with $1<q/m<\sqrt{2}$.}
	
	\section{Discussion: Cosmic Censorship as a Guiding Principle for Hawking Radiation Models}
	
	
	In this work we have examined in details the Kim-Wen model that imposed maximum mutual information principle on the tunneling picture of Hawking evaporation, in which the Reissner-Nordstr\"om black hole evolves towards extremality. Since there are a few arguments (both classical and semi-classical ones) which indicate that extremal black holes are effectively singular, this would mean that the end state of Hawking evaporation would violate cosmic censorship. Note that problems can arise even without extremality being attained exactly if the realistic extremal black hole has a singular horizon as indicated by the GHS solution after taking into account stringy correction, since curvature that is arbitrarily large is equally bad from the point of view of cosmic censorship. To make matters worse, though ordinarily the third law of thermodynamics implies that it will take an extremely long time to even come anywhere close to extremality (since the colder the black hole, the less likely it is to emit particles), this is not necessarily the case in the Kim-Wen model because particle emission is governed by the MMI selection rule and not thermodynamics per se. Therefore, there is a real danger that extremality can actually be attained. 
	
	However, one might argue that the original Kim-Wen model is not very realistic anyway since it assumes there exists a continuous set of particles with arbitrary charge-to-mass ratio. Perhaps in a more refined model with only a discrete set of charged particles the model would behave in a different manner? We showed that while the refined model does exhibit a richer behavior depending on the available particle species, \emph{generically} all end states become extremal or super-extremal or a charged remnant. In other words, a more realistic model actually makes the situation \emph{worse} with the presence of super-extremal states, thus robustly ruling out the Kim-Wen model. Though one might be tempted to judge on the ground of huge deviations from thermality alone that a Hawking evaporation model is problematic, here we provide an alternative physical guiding principle based on cosmic censorship: charged black holes should not evolve towards extremality or become super-extremal (naked singularity) under the proposed model. 
	
	Are there other effects we could include to save the Kim-Wen model? One possibility is to consider emission of multiple particles instead of one particle at a time. 
	The formula in Eq.\eqref{mutual info 2} can be generalized to an arbitrary number or correlated systems $x_1\ldots x_n$ by the following expression (see \cite{formula}):
	\begin{equation}
		S_{MI}(X; x_1,\ldots x_n) := S(x_1,\ldots, x_{n-1}\rvert x_n) - S(X; x_1,\ldots, x_{n-1}). \label{mutual info gen}
	\end{equation}
	In fact, by considering two-particle emission it is possible to emit particles whose charge-to-mass ratio exceeds $\sqrt{2}$, unlike the 1-particle emission. However, it is not clear if this is feasible. For example, suppose one particle is emitted at time $t_0$, and two particles are emitted at $t_1 > t_0$, what happens after that? Would four particles be emitted or just one? The former would lead to an exponential emission rate that seems highly impossible; while the latter (emission of $1,2,1,2,\cdots$ particles in a sequence) seems to require the black hole to somehow keep track of how many particles it has emitted in the previous steps, which is also unlikely. In any case it seems that further generalizations to the Kim-Wen model would be too ad hoc to be considered seriously, like an epicycle on top of a failed model. 
	
	Since numerically we can only probe the small mass regime, this means that the black holes analyzed in this work are highly quantum in nature, and thus one may wonder if cosmic censorship is still relevant. Of course questions like this is highly speculative, but there are indeed hints that cosmic censorship remains relevant at least at the semi-classical level. One of the strongest evidence comes from the study of the Hiscock-Weems model applied to the stringy GHS black hole \cite{1907.07490}, in which by imposing cosmic censorship (i.e., demanding that charged black holes do not become extremal) one could derive the correct charged particle production rate in agreement with direct QFT computation. Indeed, it is not clear whether singularities can be cured even in the full quantum theory of gravity (see \cite{2005.07032,2112.08531}). 
	The classical versions of singularity theorems require some form of energy conditions, whose validity is questionable in the regime where quantum gravitational effect becomes important. However, Wall has shown that even without energy conditions, the generalized second law implies a quantum singularity theorem \cite{1010.5513}. That is, if entropy remains relevant then the singularity theorem holds. If so, some form of cosmic censorship should also persist, and we could use it as a guiding principle to rule out models of Hawking evaporation. Although information is probably not lost in black holes, getting the information out remains a tricky issue that requires a better understanding of quantum information theory in the context of quantum black holes; not every efficient way to extract information is good from the point of view of cosmic censorship.

	\begin{acknowledgments}
		The authors wish to thank Wen-Yu Wen for useful discussion.
		YCO thanks the National Natural Science Foundation of China (No.11922508) for funding support. 
	\end{acknowledgments}
	
	\begin{appendices}
		\counterwithin{subfigure}{section}
		\renewcommand{\thefigure}{(A.\arabic{figure})}

		\section{Specific Heat of the Kim-Wen Model}\label{appendixa}

		\begin{figure}[h!]
			\centering
			\includegraphics[scale=0.72]{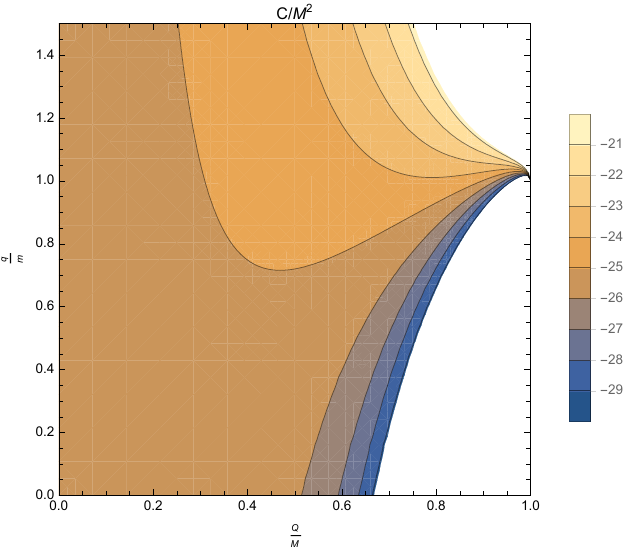}
			\caption{Specific heat in the Kim-Wen model, as a function of the ratios $Q/M$ and $q/m$.}
			\label{fig:specific heat}
		\end{figure}
	\end{appendices}
	
	Although not important for our work, it may be useful to explicitly show the details about the specific heat in this model.
	In the work of Kim and Wen \cite{1311.1656}, the specific heat can be calculated in the following way:
	\begin{equation}
		\frac{\d M}{\d T} = \left[\frac{\partial T}{\partial M}+\frac{\partial T}{\partial Q}\frac{\d Q}{\d M}\right]^{-1}.
	\end{equation}
	The variation of mass and charge is $\delta M = m, \delta Q = q$ and the ratio $q/m$ is taken to be $\gamma$ (defined in Eq.\eqref{gamma}), thus giving the result:
	\begin{equation}
		C := \frac{\d M}{\d T} \approx \frac{\delta M}{\delta T}  = 2\pi r_+^2\left[\frac{M(r_+ - 3M)+Q^2}{M^2-Q^2}\right],
	\end{equation}
	with $r_+ = M+\sqrt{M^2-Q^2}$ being the radius of the outer horizon. As shown in \cite{1311.1656}, this expression is always negative.
	To make it more general and apply it in our model, we keep the ratio $q/m$ as a variable. We then get the following expression for the specific heat:
	\begin{equation}
		C =\frac{2 \pi r_+^3 \left(r_+-M\right)}{2 Q^2 - M r_+ - Q  \left(2 M - r_+\right)\frac{q}{m}}.
	\end{equation}
	From this we can draw a contour plot of $C/M^2$ (Figure  \ref{fig:specific heat}) and confirm the negativity of the specific heat, even in the case of our discrete model, when $q/m$ does not necessarily correspond to $\gamma$.

	Kim and Wen remarked that the negative specific heat is consistent with the black hole evolution in their continuous model, namely that charged black holes tend towards extremality monotonically. This is in contrast with the Hiscock-Weems model in which the black hole specific heat changes sign, which seems to correspond to part of the attractor behavior \cite{HW}. 

\end{document}